\documentstyle[spie]{article}

\input{psfig}   
\def\jnl@style{\it}
\def\aaref@jnl#1{{\jnl@style#1}}

\def\aaref@jnl#1{{\jnl@style#1}}

\def\aj{\aaref@jnl{AJ}}                   
\def\araa{\aaref@jnl{ARA\&A}}             
\def\apj{\aaref@jnl{ApJ}}                 
\def\apjl{\aaref@jnl{ApJ}}                
\def\apjs{\aaref@jnl{ApJS}}               
\def\ao{\aaref@jnl{Appl.~Opt.}}           
\def\apss{\aaref@jnl{Ap\&SS}}             
\def\aap{\aaref@jnl{A\&A}}                
\def\aapr{\aaref@jnl{A\&A~Rev.}}          
\def\aaps{\aaref@jnl{A\&AS}}              
\def\azh{\aaref@jnl{AZh}}                 
\def\baas{\aaref@jnl{BAAS}}               
\def\jrasc{\aaref@jnl{JRASC}}             
\def\memras{\aaref@jnl{MmRAS}}            
\def\mnras{\aaref@jnl{MNRAS}}             
\def\pra{\aaref@jnl{Phys.~Rev.~A}}        
\def\prb{\aaref@jnl{Phys.~Rev.~B}}        
\def\prc{\aaref@jnl{Phys.~Rev.~C}}        
\def\prd{\aaref@jnl{Phys.~Rev.~D}}        
\def\pre{\aaref@jnl{Phys.~Rev.~E}}        
\def\prl{\aaref@jnl{Phys.~Rev.~Lett.}}    
\def\pasp{\aaref@jnl{PASP}}               
\def\pasj{\aaref@jnl{PASJ}}               
\def\qjras{\aaref@jnl{QJRAS}}             
\def\skytel{\aaref@jnl{S\&T}}             
\def\solphys{\aaref@jnl{Sol.~Phys.}}      
\def\sovast{\aaref@jnl{Soviet~Ast.}}      
\def\ssr{\aaref@jnl{Space~Sci.~Rev.}}     
\def\zap{\aaref@jnl{ZAp}}                 
\def\nat{\aaref@jnl{Nature}}              
\def\iaucirc{\aaref@jnl{IAU~Circ.}}       
\def\aplett{\aaref@jnl{Astrophys.~Lett.}} 
\def\apspr{\aaref@jnl{Astrophys.~Space~Phys.~Res.}}
\def\bain{\aaref@jnl{Bull.~Astron.~Inst.~Netherlands}} 
\def\fcp{\aaref@jnl{Fund.~Cosmic~Phys.}}  
\def\gca{\aaref@jnl{Geochim.~Cosmochim.~Acta}}   
\def\grl{\aaref@jnl{Geophys.~Res.~Lett.}} 
\def\jcp{\aaref@jnl{J.~Chem.~Phys.}}      
\def\jgr{\aaref@jnl{J.~Geophys.~Res.}}    
\def\jqsrt{\aaref@jnl{J.~Quant.~Spec.~Radiat.~Transf.}}
\def\memsai{\aaref@jnl{Mem.~Soc.~Astron.~Italiana}}
\def\nphysa{\aaref@jnl{Nucl.~Phys.~A}}   
\def\physrep{\aaref@jnl{Phys.~Rep.}}   
\def\physscr{\aaref@jnl{Phys.~Scr}}   
\def\planss{\aaref@jnl{Planet.~Space~Sci.}}   
\def\procspie{\aaref@jnl{Proc.~SPIE}}   

\makeatletter

\def\lsim{\compoundrel<\over\sim}
\def\compoundrel#1\over#2{\mathpalette\compoundreL{{#1}\over{#2}}}
\def\compoundreL#1#2{\compoundREL#1#2}
\def\compoundREL#1#2\over#3{\mathrel{\vcenter{\hbox{$\m@th\buildrel{#1#2}\over{#1#3}$}}}}
\makeatother

\title{Cross--matching DENIS and 2MASS Point Sources towards the Magellanic Clouds} 

\author{N. Delmotte\supit{a}, D. Egret\supit{a}, C. Loup\supit{b}, and M.R. Cioni\supit{c} 
\skiplinehalf 
\supit{a}CDS, Observatoire Astronomique de Strasbourg, UMR 7550, 11 rue de l'Universit\'e, \\ 67000 Strasbourg, France 
\\
\supit{b}Institut d'Astrophysique de Paris, CNRS, 98 bis Bd Arago, 75014 Paris, France
\\
\supit{c}Leiden Observatory, University of Leiden, P.O. Box 9513, 2300 RA Leiden, The Netherlands
}

\authorinfo{Further author information: (Send correspondence to N.D.)\\N.D.: E-mail: delmotte@astro.u-strasbg.fr\\ 
D.E.: E-mail: egret@astro.u-strasbg.fr\\ 
C.L.: E-mail: loup@iap.fr\\ 
M.R.C.: E-mail: mrcioni@strw.leidenuniv.nl}

\begin{document} 
  \maketitle 

\begin{abstract}
The recent publications of the DENIS Catalogue towards the
Magellanic Clouds (MCs) with more than 1.3 million sources identified in at least two
of the three DENIS filters (I J $\textrm{K}_\textrm{s}$) and of the incremental releases of the 2MASS
point source catalogues (J H $\textrm{K}_\textrm{s}$) covering the same region of the
sky, provide an unprecedented wealth of data related to stellar
populations in the MCs.
In order to build a reference catalogue of stars 
towards the Magellanic Clouds, we have performed a  cross--identification
of these two catalogues. This implied developing new tools for cross--identification and 
data mining.
 This study is partly supported by the Astrovirtel program that
aims at improving access to astronomical archives as virtual telescopes.
The main goal of the present study is to validate new cross--matching 
procedures for very large catalogues, and to derive results concerning the 
astrometric and photometric accuracy of these catalogues. The cross--matching of 
large surveys is an essential tool to improve our understanding of their 
specific contents. 
This approach can be considered as 
a new step towards a Virtual Observatory.

\end{abstract}




\section{INTRODUCTION}
\label{sect:intro}  

\begin{figure}
\begin{center}
   \begin{tabular}{cc}
   LMC--DCMC & LMC--2MASS\\
   \psfig{figure=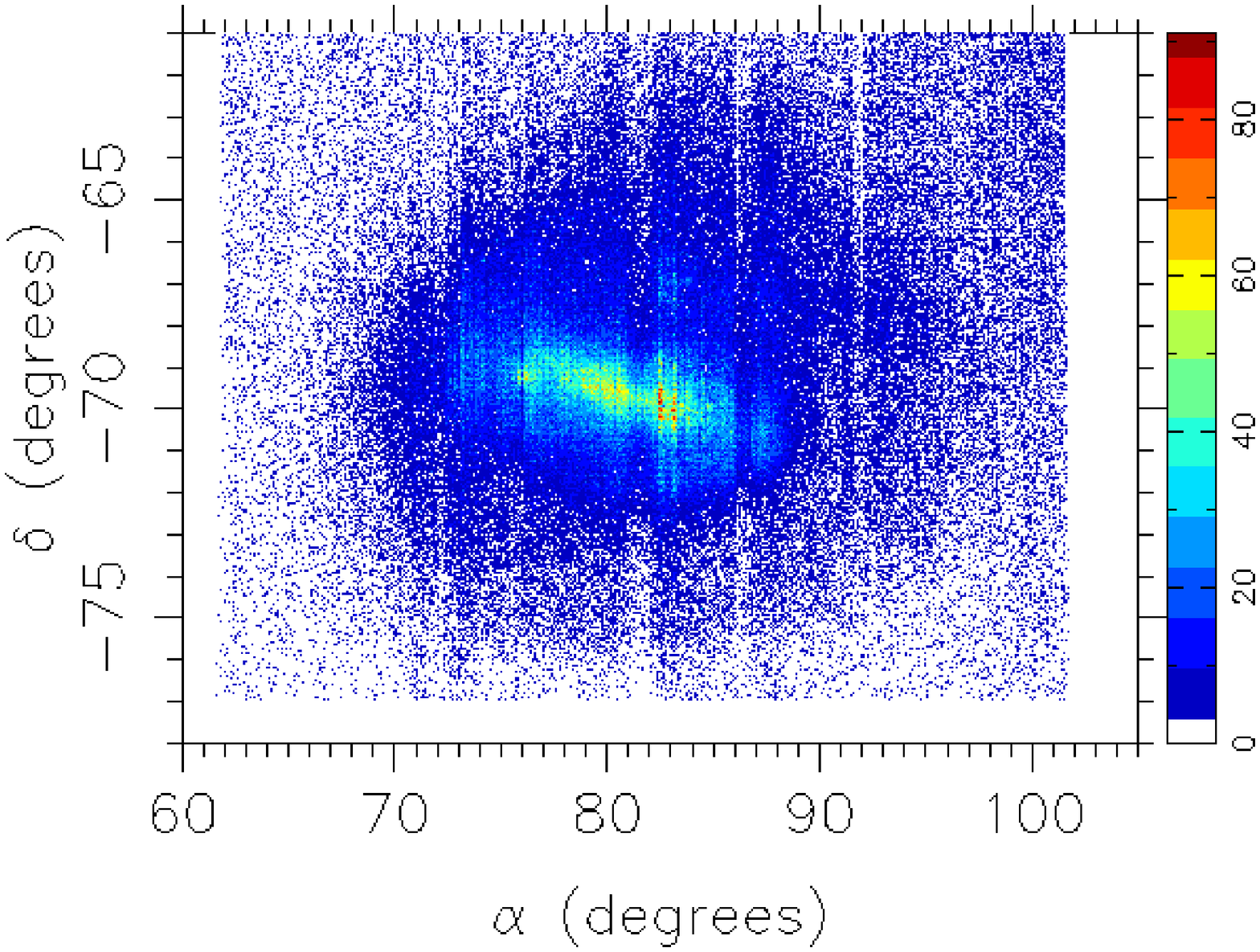,clip=,height=5cm,angle=-90} & \psfig{figure=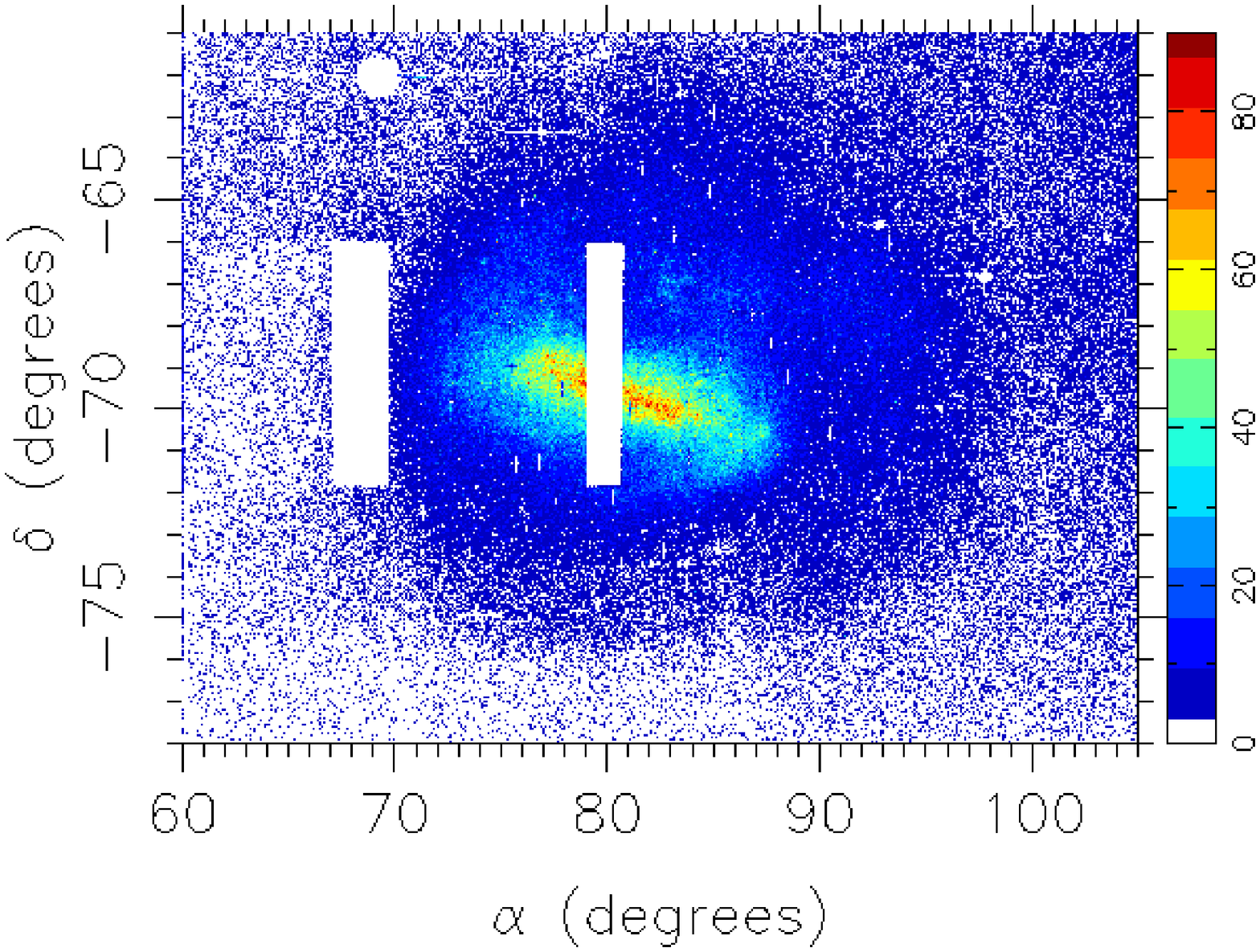,clip=,height=5cm,angle=-90} \\
   SMC--DCMC & SMC--2MASS\\
   \psfig{figure=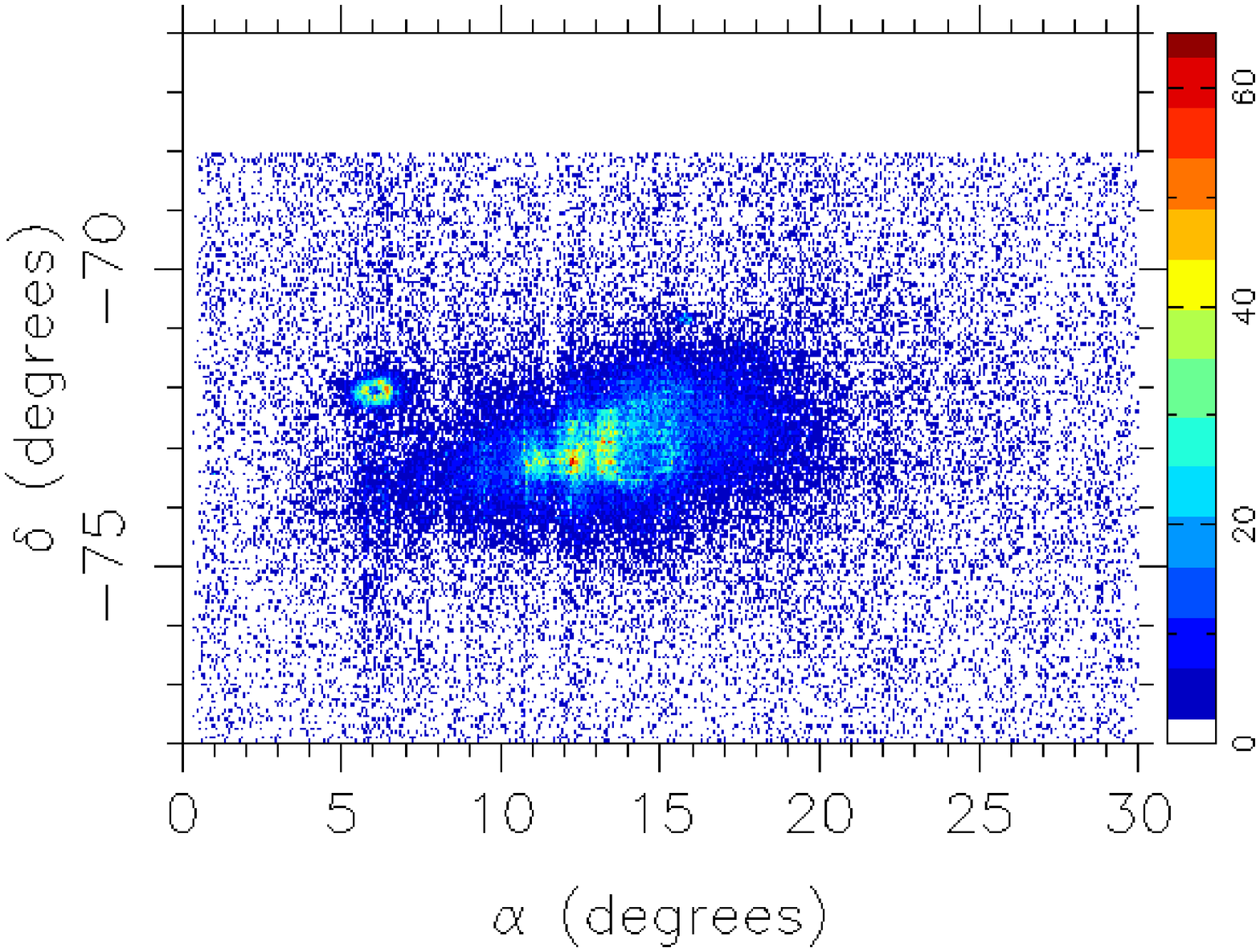,clip=,height=5cm,angle=-90} & \psfig{figure=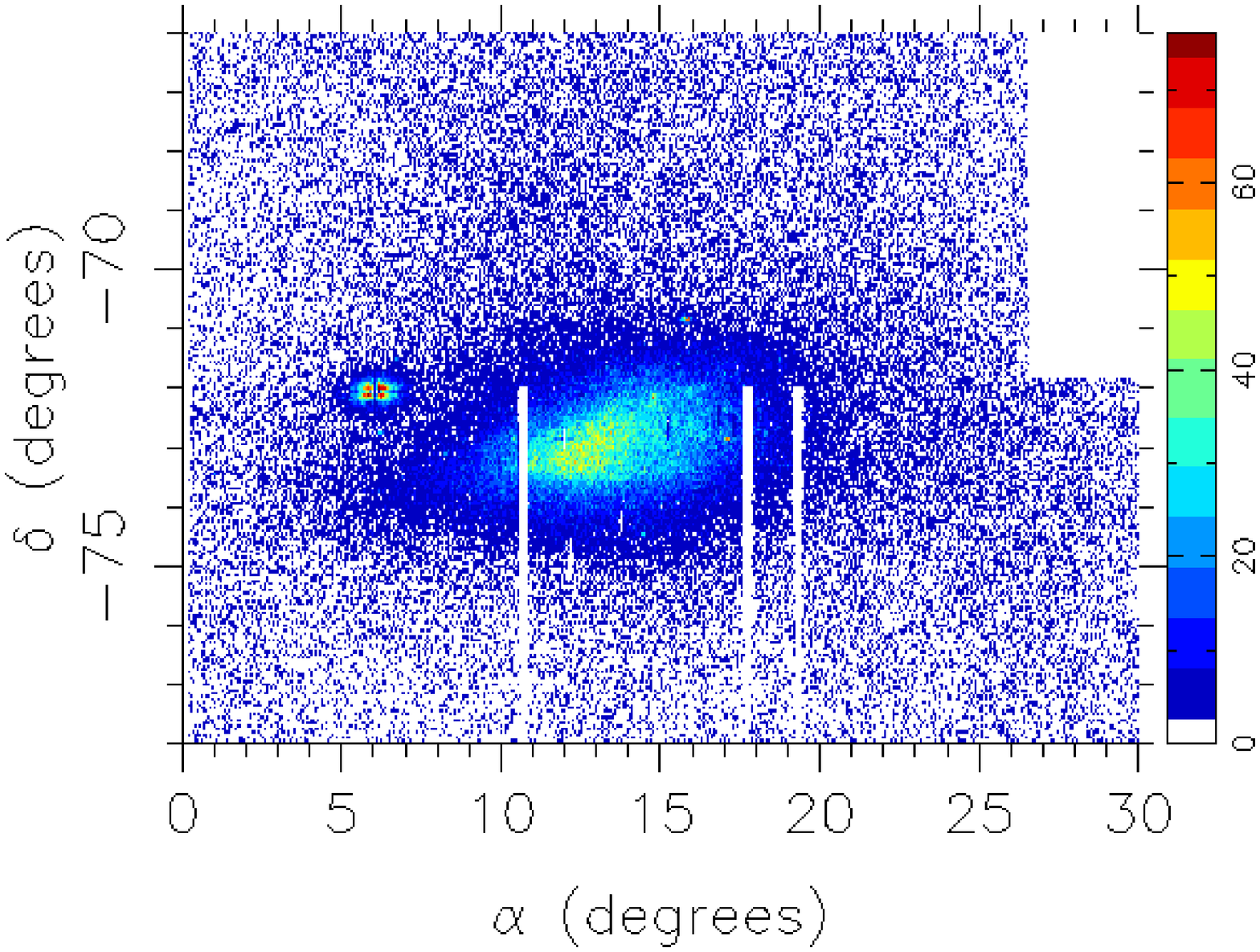,clip=,height=5cm,angle=-90} 
   \end{tabular}
   \end{center}
   \caption[example]
   { \label{fig:map}	  
 Density maps of the Clouds with respectively DCMC and 2MASS sources. The pixel size has been degraded to $3^{\prime} \times 3^{\prime}$.
 Inhomogeneities in the magnitude limit 
 of some DENIS strips is visible in the DCMC maps (left side). 
 White areas in the 2MASS maps (right side) denote missing data due to observations not yet released.} 
   \end{figure}

The Magellanic Clouds (MCs) have been recently fully observed by two major infrared surveys :
 the Deep Near Infrared Survey of the Southern Sky - DENIS\cite{Epchtein99}  and
 the Two Micron All Sky Survey - 2MASS\cite{Skrutskie97}.
A Near Infrared Point Source Catalogue towards the Magellanic Clouds, based on DENIS data,
has been published (hereafter DCMC\cite{Coco}). 
The catalogue covers an area of $19.87 \times 16$ degrees
centered on the Large Magellanic Cloud (LMC), and an area of
$14.7 \times 10$ degrees for the Small Magellanic Cloud (SMC).
To compute this catalogue, the objects were required to be detected in at least 
two of the three DENIS bands $\textrm{I} (Gunn-i, 0.79\mu m)$, $\textrm{J} (1.22\mu m)$, $\textrm{K}_\textrm{s} (2.15\mu m)$.
The 2MASS observed the whole Magellanic Clouds in three photometric bands :
$\textrm{J} (1.23\mu m)$, $\textrm{H} (1.63\mu m)$ and $\textrm{K}_{\textrm{s}} (2.15\mu m)$.
Most of the data are available from the Second Incremental Release PSC\cite{IDR2}, except for small gaps 
in regions crossing the LMC and SMC bars
and around bright stars (Fig.~\ref{fig:map}).

\begin{figure}
\begin{center}
   \begin{tabular}{c}
   \psfig{figure=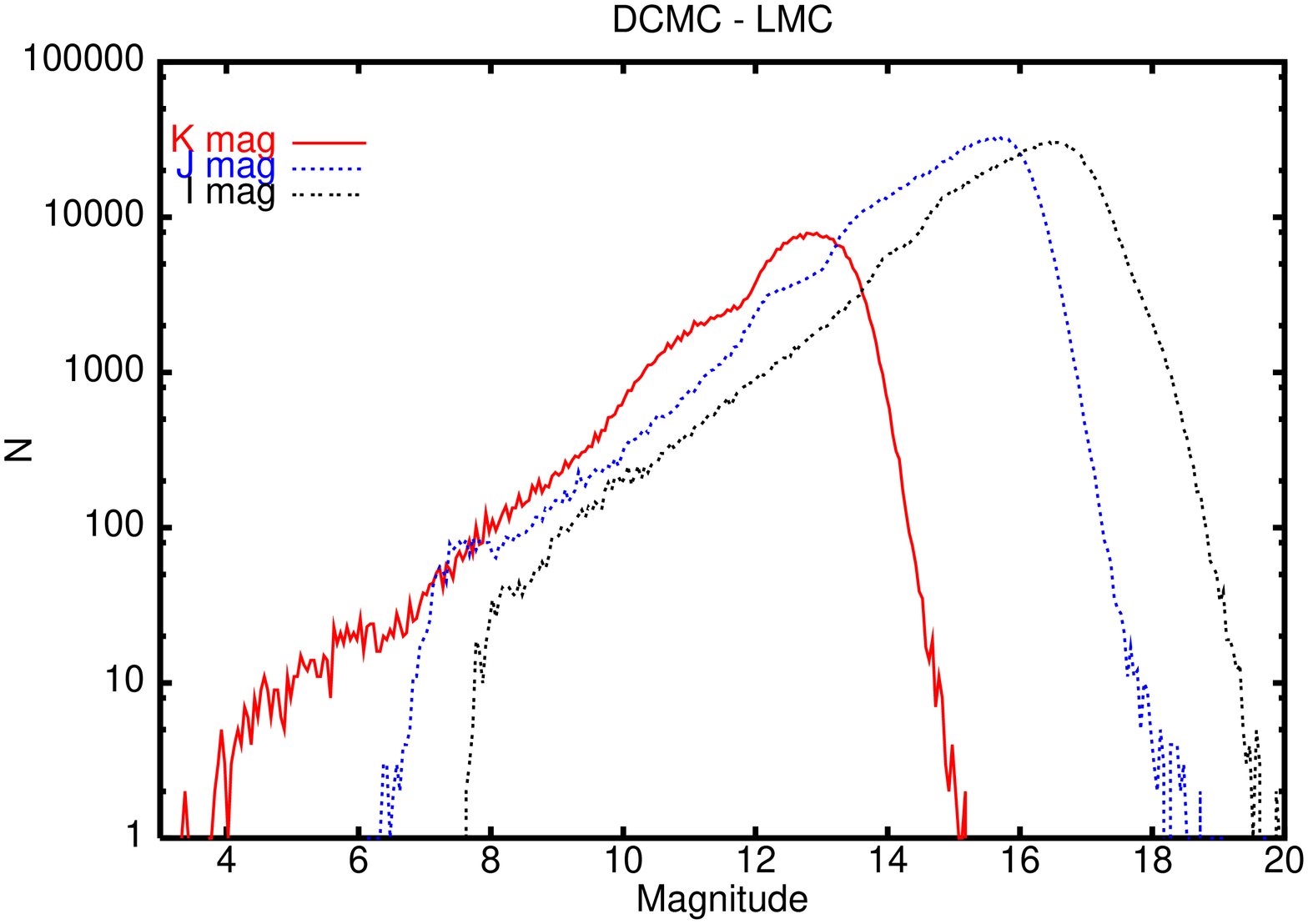,clip=,height=5cm,angle=-90} \psfig{figure=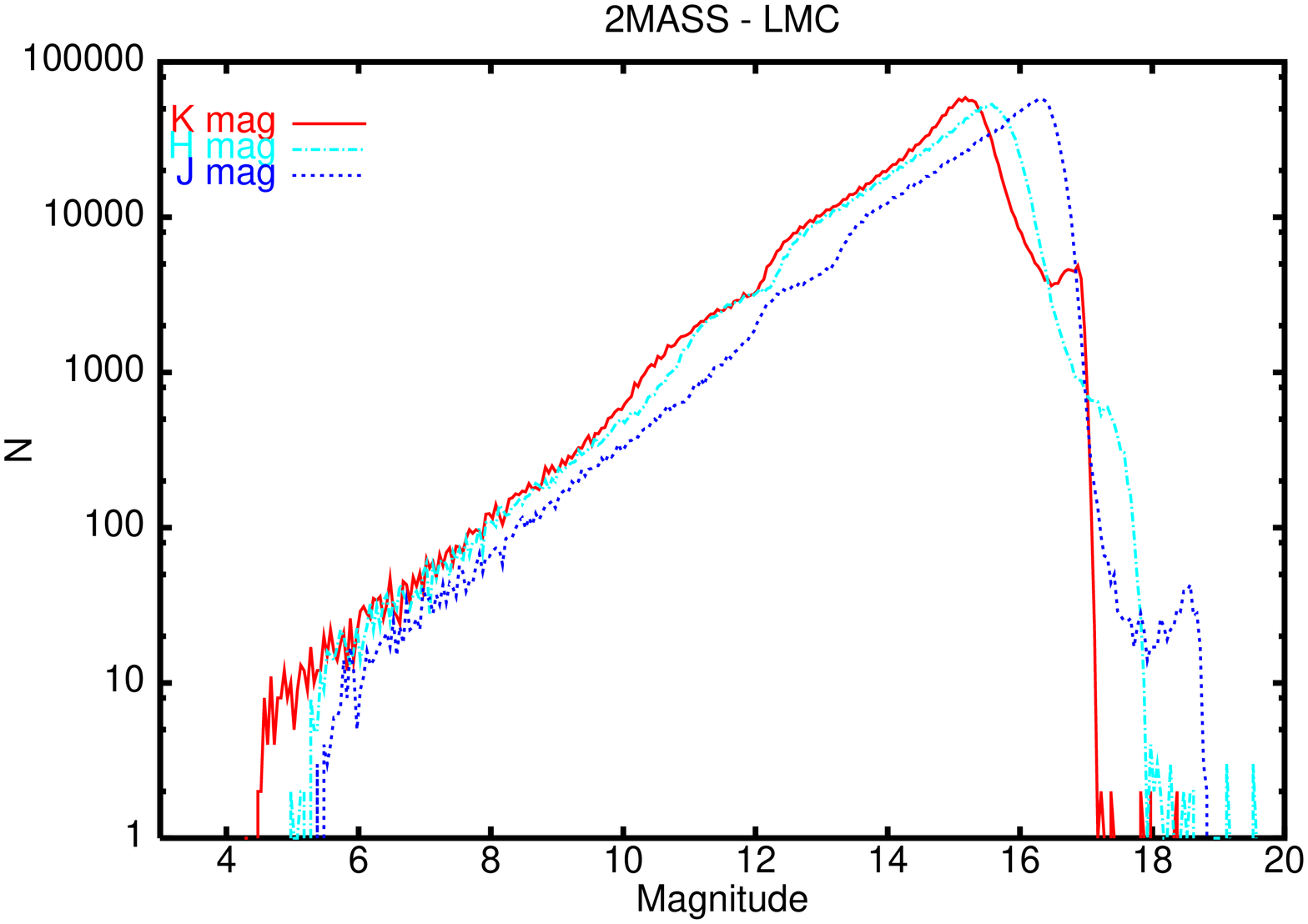,clip=,height=5cm,angle=-90} 
   \end{tabular}
   \end{center}
   \caption[example] 
   { \label{fig:compl}	 
Completeness diagrams for the LMC. 2MASS observations are deeper than DENIS ones in the $\textrm{K}_{\textrm{s}}$ band.
}
   \end{figure}

The Magellanic Clouds are one of the best places to study stellar evolution because of their proximity
and common distance of their constituent objects.
Near infrared surveys provide interesting data for this kind of 
study because of their insensitivity to dust reddening.
The number of sources from both surveys are recorded in Table~\ref{tab:nbsource}. 
Because of different sensitivity limits, DENIS sources detected only in the I and J bands are often detected
in H and $\textrm{K}_\textrm{s}$ by 2MASS.
The 2MASS observations reach almost one magnitude fainter than DENIS in the $\textrm{K}_\textrm{s}$ channel, while they
are roughly equivalent in the J channel (Fig.~\ref{fig:compl}). So it would be 
interesting to cross--match the two catalogues to complete the spectral range of the DCMC IJ--sources with the H and $\textrm{K}_\textrm{s}$ bands coming
from 2MASS.
Thus, cross--identification of the DCMC and 2MASS
catalogues will provide an unprecedented basis for study of stellar populations in the
Magellanic Clouds and for further cross--identifications with catalogues at other wavelengths. 
Furthermore the Clouds are a good place to develop and 
test cross--matching procedures for dense and large regions of the sky.

This work, partly supported by the Astrovirtel\cite{Astrovirtel} program,
 aims at performing a systematic cross--matching of the DCMC with 2MASS,
and a series of validation tests that will enable us to  understand better the specific contents of
each catalogue.

\begin{table} [h]   
\label{tab:nbsource}
\begin{center}
\caption{Number of sources as a function of detected wavebands. {\em None} means saturated sources.}
\begin{minipage}[r]{0.49\linewidth}
	\centering
	\begin{tabular}{rr|rr}
	\multicolumn{4}{c}{\bf LMC}  \\
	\hline
	\hline
	\multicolumn{2}{c|}{DCMC} & \multicolumn{2}{c}{2MASS}\\
	\hline
	IJ$\textrm{K}_\textrm{s}$ & 297031 & JH$\textrm{K}_\textrm{s}$ & 1996382 \\
	IJ & 1151789 & J$\textrm{K}_\textrm{s}$ & 66 \\
	I$\textrm{K}_\textrm{s}$ & 8724 & JH &  - \\
	J$\textrm{K}_\textrm{s}$ & 1897 & H$\textrm{K}_\textrm{s}$ & 4 \\
	 &  & J & 11 \\
	 &  & H & - \\
	 &  & $\textrm{K}_\textrm{s}$ & 23 \\
	 &  & None & 259 \\
	\hline 
	Total & 1459441 & Total & 1996745 \\
	\end{tabular}
\end{minipage}
\hfill
\begin{minipage}[l]{0.49\linewidth}
	\centering
	\begin{tabular}{rr|rr}
	\multicolumn{4}{c}{\bf SMC}   \\
	\hline
	\hline
	\multicolumn{2}{c|}{DCMC} & \multicolumn{2}{c}{2MASS}\\
	\hline
	IJ$\textrm{K}_\textrm{s}$ & 75133 & JH$\textrm{K}_\textrm{s}$ & 481549 \\
	IJ & 257925 & J$\textrm{K}_\textrm{s}$ & 15 \\
	I$\textrm{K}_\textrm{s}$ & 2206 & JH &  1 \\
	J$\textrm{K}_\textrm{s}$ & 710 & H$\textrm{K}_\textrm{s}$ & 4 \\
	 &  & J & 4 \\
	 &  & H & - \\
	 &  & $\textrm{K}_\textrm{s}$ & 9 \\
	 &  & None & 69 \\
	\hline 
	Total & 335974 & Total & 481651 \\
	\end{tabular}
\end{minipage}
\end{center}
\end{table}

\section{DATA ORGANIZATION} 


The present work is based on public data from 2MASS, as given in Sect. \ref{sect:intro}.
DCMC data have been obtained from a local copy of the catalogue that includes the missing strips
of the first release (the second release is currently under process). 
Before running the cross--matching programs, we organized the raw data,
splitting both catalogues into smaller pieces.
The DENIS observational strategy has been to divide the sky in strips of $30^{\circ}$ in Declination (DEC)
and $12^{\prime}$ in Right Ascension (RA).
To define subsamples, we adopted a strip by strip strategy because :
\begin{itemize}
\item Our cross--matching algorithm is well adapted to data files with small extension in RA (alpha) and
a strip is only $12^{\prime}$ large in RA.
\item The cross--matching criteria depend on the strip number as explained below (see Sect. \ref{sect:cross}). 
\end{itemize} 
First we split the DCMC catalogue by strip number. 
There are 119 strip--files for the LMC and 88 strip--files for the SMC.
Then for each strip--file, we extracted from the 2MASS data all the point sources overlapping the same region of the sky.

The cross--matching program is run for each strip. Each time we have two input 
files, one from DCMC, one from 2MASS, corresponding to a given strip number.  
Both files have been previously sorted by ascending declination, in order to optimize 
the cross--comparison procedure. 
The procedure can be described as follows : both files are read sequentially in 
parallel ; for each record of the first file (say, DCMC) we search for all possible 
cross--matches in the second file (here 2MASS). For that, we read the second file 
and keep in memory a buffer of possible candidates, making sure that the highest 
value of the declination in file 2 is actually higher than $\delta$ + $\delta_{\delta}$ 
of the current record of file 1, and the same thing for the lowest value of the buffer 
of file 2, which has to be lower than $\delta$ - $\delta_{\delta}$.
Possible cross--matches are kept, together with the corresponding differences in 
positions and magnitudes. In a first run, we keep the smallest difference in position 
as the most probable, using a box of $\delta_{\alpha}$, $\delta_{\delta}$ = $10^{\prime\prime}$.

\section{FIRST CROSS--MATCHING STEP : FINDING DISCREPANCIES IN THE ORIGINAL CATALOGUES}

\begin{figure}
\begin{center}
   \begin{tabular}{c}
   \psfig{figure=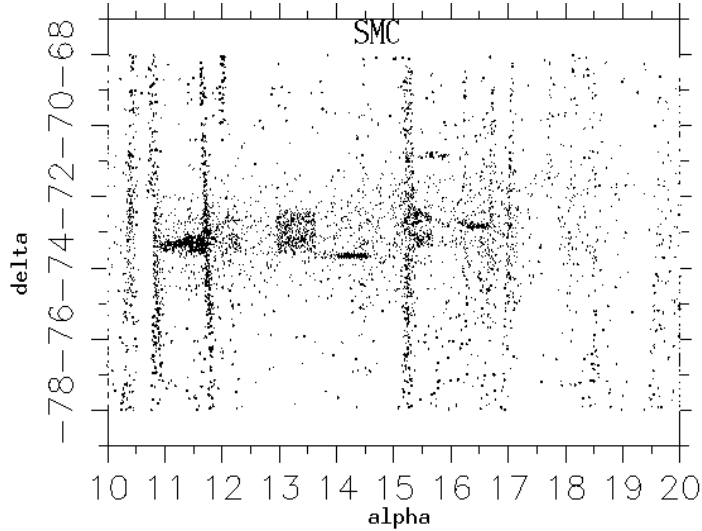,clip=,height=7cm} 
   \end{tabular}
   \end{center}
   \caption[example] 
   { \label{fig:colordot}	 
$\alpha$ and $\delta$ are in degrees. They span the central part of the SMC.
 Each dot corresponds to one cross--identification with a distance larger than $1^{\prime\prime}$.
 The geometric patterns indicate that subsets (images and strip overlaps) of the DCMC present errors in their astrometric calibration. }
   \end{figure}

The easiest way to find matches between two catalogues is to fix a searching box in position of a few arcseconds and compare
the coordinates. It will work really well in most cases because the astrometry
of the two DCMC and 2MASS catalogues is accurate enough (better than one arcsecond). Furthermore both catalogues were calibrated upon
the USNO-A2.0 catalogue\cite{Monet98}. Consequently the distance match is better than $0.5^{\prime\prime}$ for the great majority of stars. There is in principle no risk of
confusion at such a small scale. While this is true in general, in practice the cross--matching exercice has proven to be a 
powerful tool to detect subsets of the data files which deviate from the perfect situation, and primarily areas suffering from problems
in the
astrometric or photometric calibration.
Here is what we did to find out these regions :
\begin{itemize}
\item We split each catalogue into big chunks : $5^{\circ}$ width in RA for the LMC and $10^{\circ}$ width in RA for the SMC. 
\item     We ran a cross--matching program based on distances {\em only}. That is for one DCMC source, we searched in 2MASS
     for all the possible matching sources in a radius of $10^{\prime\prime}$. 
\item     Between all the possible associations found, we kept only the association with the closest distance. 
\item    Then we made a map in $[\alpha,\delta]$ of the distances to the closest neighbours : 
\begin{itemize}        
	\item   Each cross--identification is marked with a dot in the $[\alpha,\delta]$ plane. 
	\item   The color of each dot depends on the distance of the cross--identification. 
\end{itemize}
\end{itemize}
Fig.~\ref{fig:colordot} shows the distance map for the central part of the SMC. We plotted only the cross--identifications with
distances larger than $1^{\prime\prime}$. They cannot be taken for random associations because geometrical and well-defined patterns 
appear on the map. 
These patterns reveal problems in the astrometry
for a few DCMC images and along the border of several strips. 
We found two main reasons to explain these results : systematic distance shifts associated with redundant DCMC sources and
non-systematic effects dealing with field distortions. 

\subsection{Redundant Sources}

\begin{figure}
\begin{center}
   \begin{tabular}{c}
   \psfig{figure=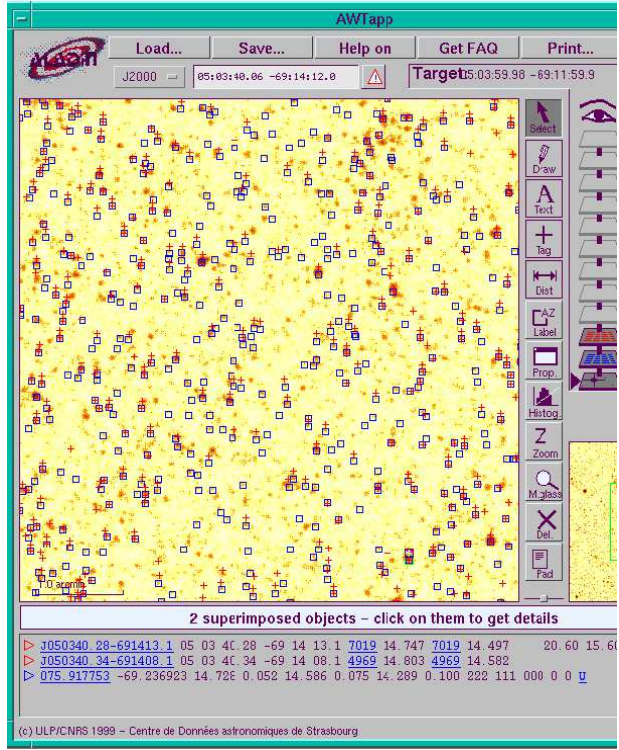,height=10cm} 
   \end{tabular}
   \end{center}
   \caption[example] 
   { \label{fig:aladin}	  
Aladin view (MAMA scan of an ESO plate) of a LMC region containing redundant DCMC sources shifted by $5^{\prime\prime}$
in declination (crosses). 2MASS entries are plotted as squares. These redundant sources are always located on the
overlaps between adjacent images and strips.} 
   \end{figure}

Redundant sources are located on the overlaps with adjacent strips, and with adjacent images 
of the same strip.
Fig.~\ref{fig:aladin} shows the Aladin view of a LMC region containing redundant
 DCMC sources. Aladin\cite{Bonnarel00} is an interactive software sky atlas developed by CDS, allowing one to visualize digitized
images of any part of the sky and to superimpose entries from astronomical catalogues.
Redundant DCMC sources are systematically shifted by $5^{\prime\prime}$ in declination above 2MASS sources. This problem mainly occurs
in crowded regions where the number of USNO-A2.0 reference stars is small because of the confusion. It can happen that an
astrometric reference star was incorrectly cross--identified with a DENIS source, leading to a systematic shift in RA
or/and in DEC. It usually affects only one image, sometimes a few adjacent images. The DENIS sources located in the overlaps
with the adjacent images of the same and adjacent strips were thus not properly
cross--identified with the DENIS sources of those adjacent images.	

\begin{figure}
\begin{center}
   \begin{tabular}{c}
  \psfig{figure=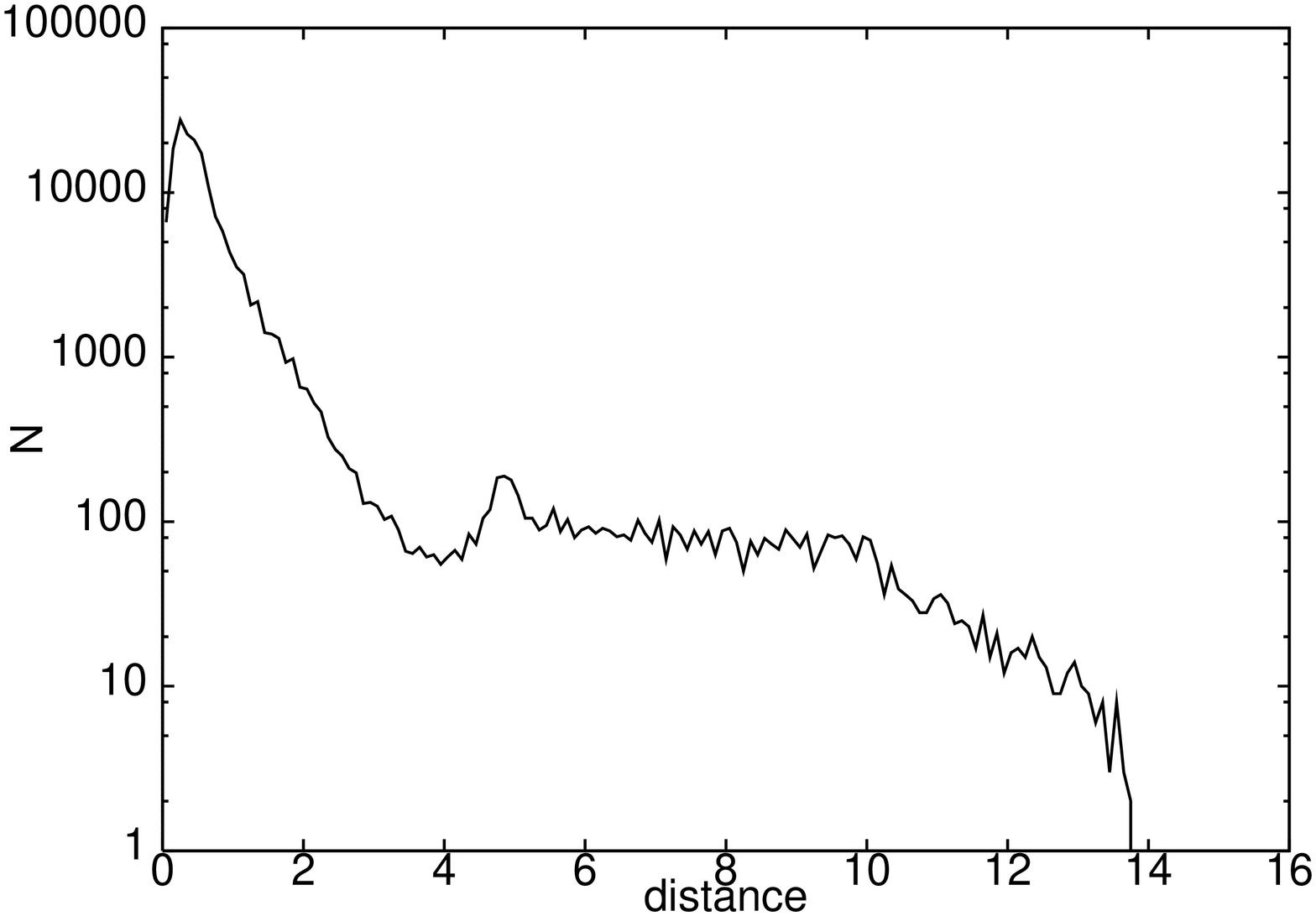,height=5cm,angle=-90}  \psfig{figure=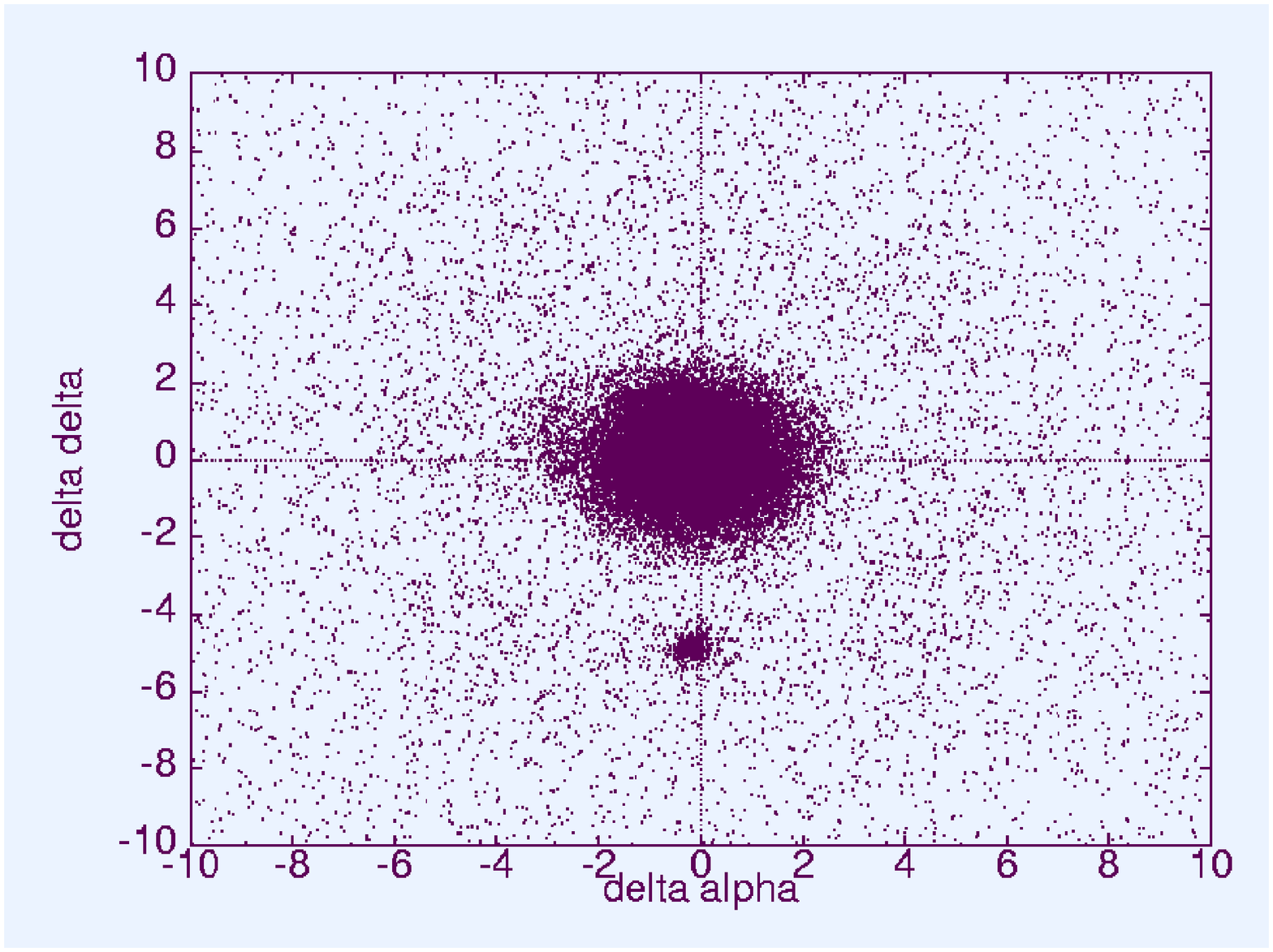,height=5cm,angle=-90}\\
    \psfig{figure=bad.ps2,height=5cm,angle=-90} 
   \end{tabular}
   \end{center}
   \caption[example] 
   { \label{fig:dist}	  
(a)~(Top left)~Histogram of distances in arcseconds between DCMC and 2MASS matched sources.
(b)~(Top right)~Each cross--matched source is marked with a dot. $\delta \alpha$ and $\delta \delta$ are in arcseconds. 
(c)~(Bottom)~$\alpha$ and $\delta$ are in degrees. Each point corresponds to one cross--identification with $4.5^{\prime\prime} < \textrm{distance} < 5.5^{\prime\prime}$.}
\end{figure}

To find out the consequences of these redundant sources on the cross--matching, we took an area of 3 x 2.7 
degrees in the LMC with $-71^{\circ} < \delta < -68^{\circ}$ and $75^{\circ} < \alpha < 82.5^{\circ}$, and including redundant sources. \\
First we made a histogram of the distances to the closest neighbours (Fig.~\ref{fig:dist} (a)). The physical associations are 
located on the left part of the histogram (distance $\lsim 2^{\prime\prime}$), whereas the non-physical random associations are on the right part.
This general feature is complemented here by
a rather striking effect : a bump is clearly visible around~$5^{\prime\prime}$.\\
Then to understand better what was going on, we considered $\alpha$ and $\delta$ separately (Fig.~\ref{fig:dist} (b)).
Each point represents one association between DCMC and 2MASS. 
On the x-axis, we have : 
\[
                             (\alpha_\textrm{2MASS} - \alpha_\textrm{DCMC}) \times \cos \delta_\textrm{DCMC} \]
and on the y-axis : 
\[                                       
			\delta_\textrm{2MASS} - \delta_\textrm{DCMC} .\]
			
If there is no significant shift between DCMC and 2MASS, all points should be centered
around (0,0), which is the case for the great majority of stars. But we can see another
cluster of points around (0,-5) which corresponds to the relative shift of the redundant
sources ($5^{\prime\prime}$ in $\delta$, $0^{\prime\prime}$ in $\alpha$). 
To characterize more precisely the faulty images, we took all the cross--identifications with
distances between $4.5^{\prime\prime}$ and $5.5^{\prime\prime}$ and we plotted them in $[\alpha, \delta]$ (Fig.~\ref{fig:dist} (c)). 
They appear to be well located inside a square region of the sky : the images 70 and 71 of the strip 4969.
All images affected by redundant sources were discarded for the following of the procedure.
12 images are concerned
in the LMC (0.1\%) and 42 in the SMC (0.8\%).


\subsection{Field Distortions} \label{sect:field-dist}

Field distortions in the DCMC affect the quality of the astrometry. To detect them, we proceeded {\em strip by strip} as follows :
\begin{itemize}      
   \item     We kept only well confirmed DCMC sources : $10.5 < \textrm{I} < 16.5 $ and flags in the I band equal to zero. 
   \item     We ran a cross--matching program based only on distances, with a searching box that goes
     up to $30^{\prime\prime}$. 
   \item     Between all the possible associations found, we kept only the association with 
   $  |\textrm{J}_\textrm{DCMC} - \textrm{J}_\textrm{2MASS}| < 0.5 $. The selection is done on magnitude because in case of field distortions 
   small
    distances are not a reliable enough criterion. 
\end{itemize}

\begin{figure}
\begin{center}
   \begin{tabular}{c}
   \psfig{figure=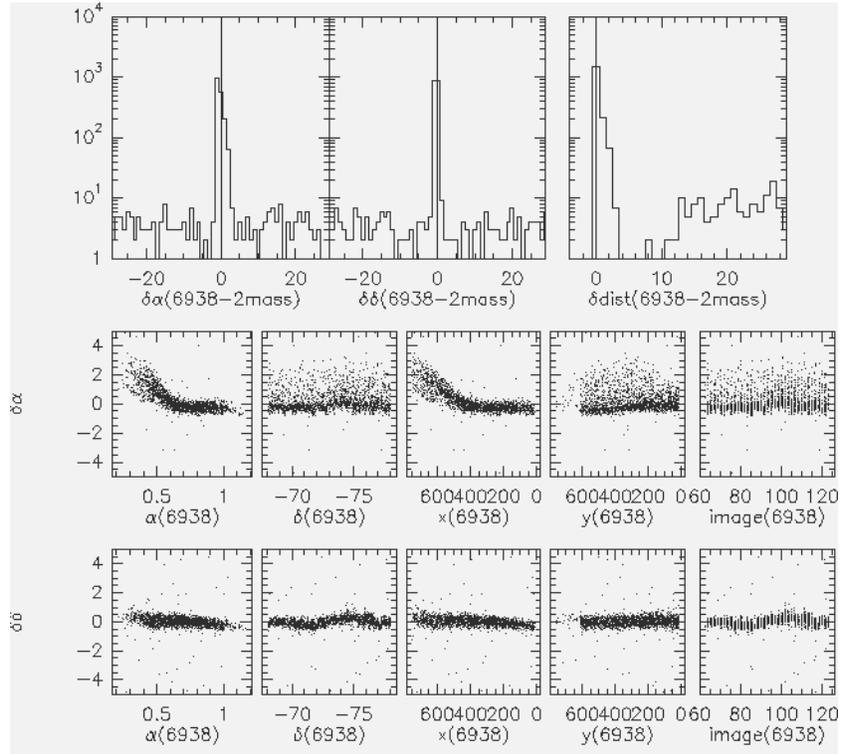,clip=,height=10cm} 
   \end{tabular}
   \end{center}
   \caption[example] 
   { \label{fig:cecile}	  
Typical example of field distortion, especially along the x-axis of the camera.
The results presented here are for the strip 6938 located in the SMC.} 
   \end{figure}

As an example, the results of the cross--identification for strip 6938 are summarized in Fig.~\ref{fig:cecile}.
The relative shifts $\delta \alpha$ and $\delta \delta$ 
are a function of the pixel coordinates (x,y) of the camera. The maximum shift in $\alpha$
between DCMC and 2MASS goes up to $3.5^{\prime\prime}$. We found 11 and 14 strips affected by field distortions at a level 
larger than $2^{\prime\prime}$ in the LMC 
and SMC, respectively. One of them (strip 5830 in the SMC) had to be rejected because of erroneous astrometric calibration.

\subsection{Photometry} \label{sect:photom}

\begin{figure}
\begin{center}
   \begin{tabular}{c}
   \psfig{figure=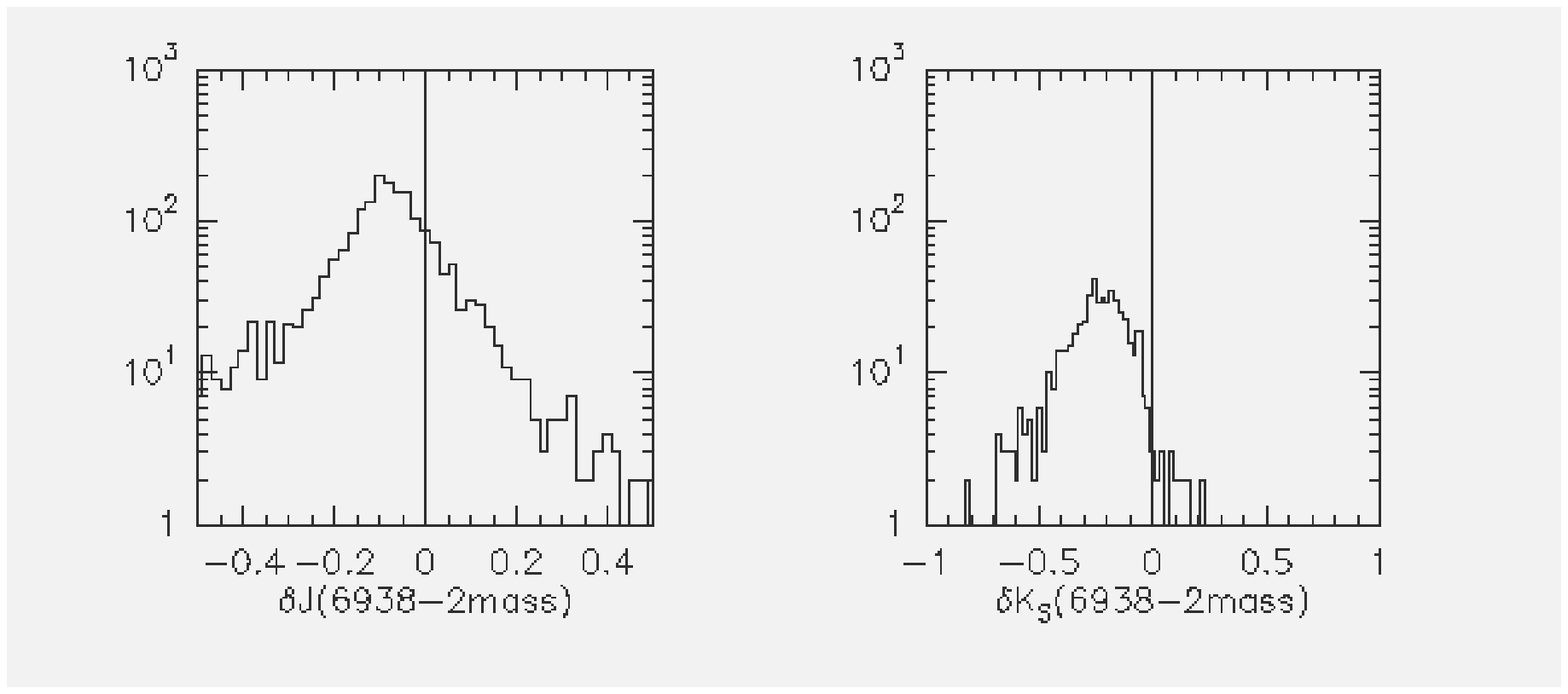,clip=,height=5cm} 
   \end{tabular}
   \end{center}
   \caption[example] 
   { \label{fig:mag}	Magnitude histograms for cross--matched sources of strip 6938.
  (Left)~$\textrm{J}_\textrm{\scriptsize DCMC}-\textrm{J}_\textrm{\scriptsize 2MASS}$ histogram. \\
(Right)~$\textrm{K}_{\textrm{s}_\textrm{DCMC}} - \textrm{K}_{\textrm{s}_\textrm{2MASS}}$~histogram.  
} 
   \end{figure}

We also searched for a systematic shift in magnitude between the DCMC and 2MASS for the J and $\textrm{K}_\textrm{s}$ bands. 
We used the cross--identifications coming from Sect. \ref{sect:field-dist}. 
Mean shifts in J and $\textrm{K}_\textrm{s}$ have been computed for {\em each} strip. The results presented on
Fig.~\ref{fig:mag} are for the strip 6938 (slot 4034). The magnitude shift is -0.07 for the J band 
and -0.24 for the $\textrm{K}_\textrm{s}$ band.

\section{CROSS--MATCHING CRITERIA} 
\label{sect:cross}
The astrometric and magnitude shifts depend on the strip and have to be taken into account. Strategies for coping with them have been implemented,
to allow a proper strip by strip cross--matching of both catalogues. For each DCMC source of a given strip, 
we search the best 2MASS association using both position and magnitude criteria :
\begin{enumerate}
\item {\bf Selection on coordinates }:
      The shifts in $\alpha$ and $\delta$ vary inside the images of the strip 
         {\em but}  
         distortions do not vary significantly along the strip. They are approximately the same for all	the images of the
         strip. So it is better to use the statistics of the whole strip instead of one single image. 
	 Thus we can define a specific searching box for the strip. The size of the box will take into account the shifts
	  in $\alpha$ and $\delta$ found for this strip number as explained in Sect. \ref{sect:field-dist}.
	  The default size of the searching box when there are no shifts is $3^{\prime \prime}$.
	  So we have now an enlarged and assymetric searching box :

       \[        \frac{\delta \alpha_\textrm{min} - 3^{\prime\prime}}{ \cos \delta} < \alpha_\textrm{DCMC} - \alpha_\textrm{2MASS} <  \frac{\delta        \alpha_\textrm{max}  +3^{\prime\prime}}{\cos \delta}
          \]           
	  \[              \delta \delta_\textrm{min}        < \delta_\textrm{DCMC} - \delta_\textrm{2MASS} < \delta  \delta_\textrm{max}	\quad  ,
	\]
	
	where $\delta \alpha_\textrm{min} $,  $\delta  \alpha_\textrm{max} $,   $\delta \delta_\textrm{min} $,  $\delta  \delta_\textrm{max}  $
	are the minimum and maximum shifts in right ascension and declination. 
\item {\bf Selection on magnitudes }:
     Between all the possible associations found in step 1, we must keep the best one. 
     We have seen that keeping the association with the smallest distance is no more a
     reliable criteria because of field distortions. So we have to check the compatibility in
     magnitude for each association, after applying on the strip data the associated mean magnitude shifts 
     $<\delta \textrm{J}>$ and $<\delta \textrm{K}_\textrm{s}>$ computed in Sect. \ref{sect:photom}.

\begin{itemize}
\item         If $\textrm{K}_\textrm{s}$ is not detected in one or both catalogue, the selection is done on J. The following 
relation has to be true to keep the association :
\begin{eqnarray*}
	|\delta \textrm{J} - <\delta \textrm{J}>|  \leq w  \times \sqrt {\sigma_{\textrm{J}_\textrm{\scriptsize DCMC}}^{2} +	\sigma_{\textrm{J}_\textrm{\scriptsize	2MASS}}^{2}}	\quad,
\end{eqnarray*} 	
where $w=2$ is a weight, and $\sigma_{\textrm{J}_\textrm{\scriptsize DCMC}}$ and $\sigma_{\textrm{J}_\textrm{\scriptsize	2MASS}}$ are the relative
photometric uncertainties as quoted in both catalogues. Relative uncertainties
are in general very small for bright stars, less than 0.01 mag. However, uncertainties on the absolute 
calibration are much larger : about 0.1 mag for the DCMC. If we apply abruptly
the above criterion, we will lose many cross--identifications for the stars with small relative uncertainties.
We thus need to refine the selection criterion and consider two cases~: 
\begin{eqnarray*}
      \textrm{if} \qquad w \times \sqrt {\sigma_{\textrm{J}_\textrm{\scriptsize DCMC}}^{2} + \sigma_{\textrm{J}_\textrm{\scriptsize           2MASS}}^{2}} \leq               \Delta      \textrm{J} & \textrm{then} & |\delta \textrm{J} - <\delta \textrm{J}>| \leq \Delta \textrm{J} \quad , \quad \textrm{else}\\
      \textrm{if} \qquad w \times \sqrt {\sigma_{\textrm{J}_\textrm{\scriptsize DCMC}}^{2} + \sigma_{\textrm{J}_\textrm{\scriptsize               2MASS}}^{2}} >        \Delta\textrm{J}             & \textrm{then} &      |\delta \textrm{J} - <\delta \textrm{J}>|  \leq w  \times \sqrt      {\sigma_{\textrm{J}_\textrm{\scriptsize DCMC}}^{2} +      \sigma_{\textrm{J}_\textrm{\scriptsize  2MASS}}^{2}} \quad, 
\end{eqnarray*} 
where  $\Delta      \textrm{J}=0.45$ is the maximum width of the $\delta \textrm{J}$ distribution.
  \item        If J is not detected in one or both catalogue, the selection is done on $\textrm{K}_\textrm{s}$ as above but this time we have  $\Delta  \textrm{K} = 0.60$.

 \item       If J and $\textrm{K}_\textrm{s}$ are detected in both catalogues, the selection is done on J and then on J-$\textrm{K}_\textrm{s}$. 


   \item       If J and $\textrm{K}_\textrm{s}$ are not detected in one or both catalogue, the association is lost. 
	 
\end{itemize}

\item Applying these criteria, if there are still more than one possible association for one DCMC source, then
     we keep the association with the smallest $\delta \textrm{J}$ or $\delta \textrm{K}_\textrm{s}$.

\end{enumerate}

\section{RESULTS}

Running the cross--matching programs took about two hours for each Cloud on a Unix station.
For each strip, we computed the percentage of DCMC sources matched with 2MASS.
Results are summarized in Table 2. Nearly 80\% of the LMC strips 
and 70\% of the SMC strips 
have a match rate better than 90\%. The 16 LMC and 18 SMC strips with a match rate smaller than 80\% correspond 
to the gaps in the 2MASS data.
The merged point sources have at least two of the four photometric bands and J or $\textrm{K}_\textrm{s}$
is always present because it was the magnitude link between the two catalogues. Table~3
lists the number of merged sources as a function of detected wavebands. When J or $\textrm{K}_\textrm{s}$
is present, it comes either from the DCMC or 2MASS, or from both. 

86\% and 83\% of the DENIS point sources
are matched with 2MASS for the LMC and SMC, respectively. The number of stars in common is 1252700 in the LMC
and 278856 in the SMC.

 \begin{table} [h]   
\label{tab:poupou}
\begin{center}
\caption{Number of strips per match rate.}
\begin{minipage}[r]{0.49\linewidth}
	\centering
	\begin{tabular}{cc|cc}
	\hline
	\hline
	\multicolumn{2}{c|}{LMC} & \multicolumn{2}{c}{SMC}\\
	\hline
	Match rate & Number of strips & Match rate & Number of strips\\
	\hline
	$> 95\%      $ & 42 & $> 95\%	  $&  30 \\
	$[90\%,95\%] $ & 52 & $[90\%,95\%] $  &  30 \\
	$[80\%,90\%] $ &  9 & $[80\%,90\%] $  &  10 \\
	$< 80\%      $ & 16 & $< 80\%	 $ &  18 \\
	\hline 
	Total & 119 & Total & 88 \\
	\end{tabular}
\end{minipage}
\end{center}
\end{table}

\begin{table} [h]   
\label{tab:merged}
\begin{center}
\caption{Merged DENIS and 2MASS point sources.}
	\centering
	\begin{tabular}{rr|rr}
	\hline
	\hline
	\multicolumn{2}{c|}{LMC} & \multicolumn{2}{c}{SMC}\\
	\hline
	IJ                          & 3                &	IJ                          & 0             \\
	I$\textrm{K}_\textrm{s}$    & 0                &	I$\textrm{K}_\textrm{s}$    & 0              \\
	J$\textrm{K}_\textrm{s}$    & 10               &	J$\textrm{K}_\textrm{s}$    & 1             \\
	IJ$\textrm{K}_\textrm{s}$   & 58               &	IJ$\textrm{K}_\textrm{s}$   & 16              \\
	IJH                         & 0                &	IJH                         & 0             \\ 
	IH$\textrm{K}_\textrm{s}$   & 0                &	IH$\textrm{K}_\textrm{s}$   & 0             \\
	JH$\textrm{K}_\textrm{s}$   & 699              &	JH$\textrm{K}_\textrm{s}$   & 185             \\
	IJH$\textrm{K}_\textrm{s}$  & 1251930          &	IJH$\textrm{K}_\textrm{s}$  & 278654             \\
	\hline 
	Total                       & 1252700          &        Total                       & 278856             \\
	\end{tabular}
\end{center}
\end{table}

      \begin{figure}
\begin{center}
   \begin{tabular}{c}
   \psfig{figure=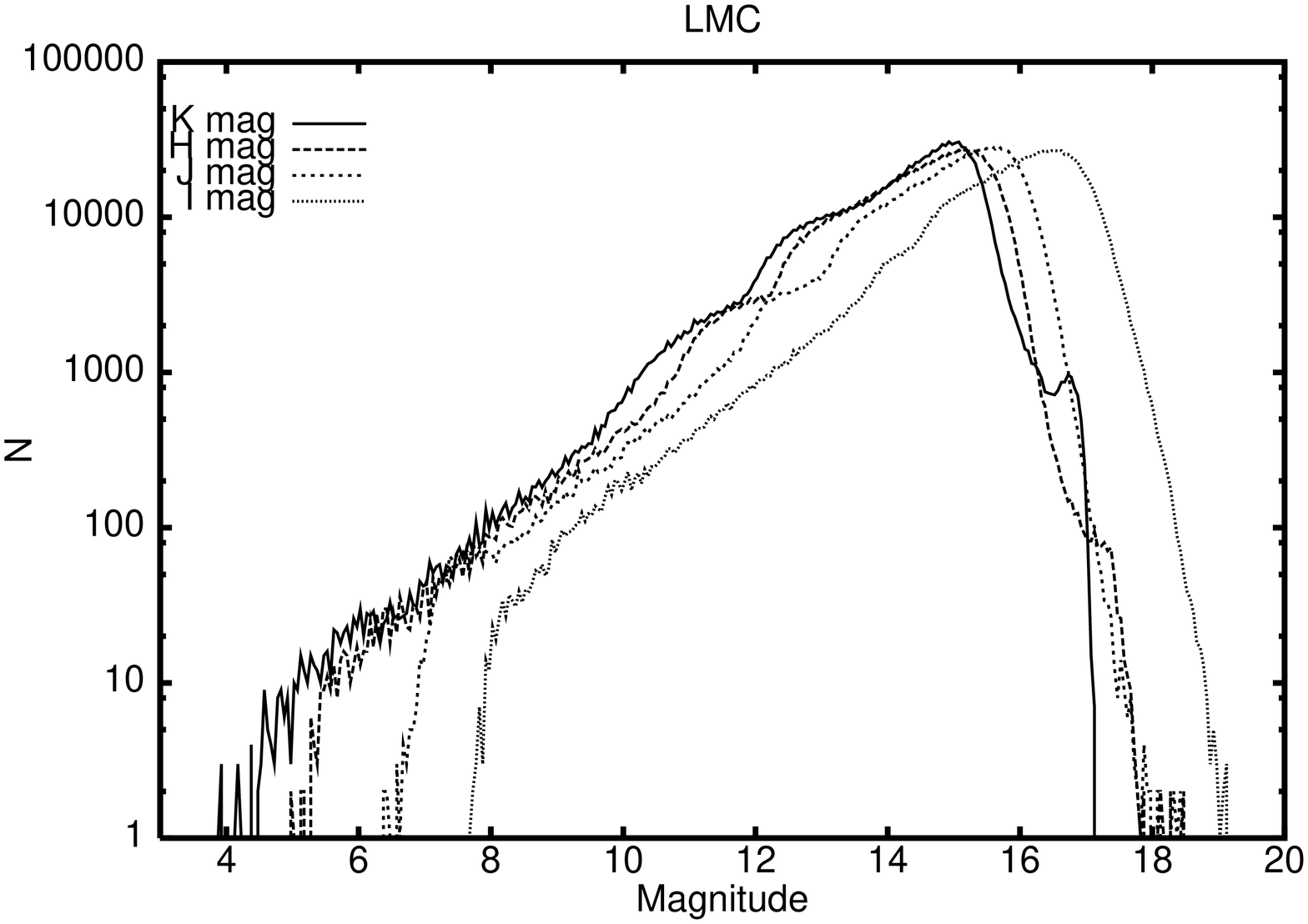,height=5cm,angle=-90} \psfig{figure=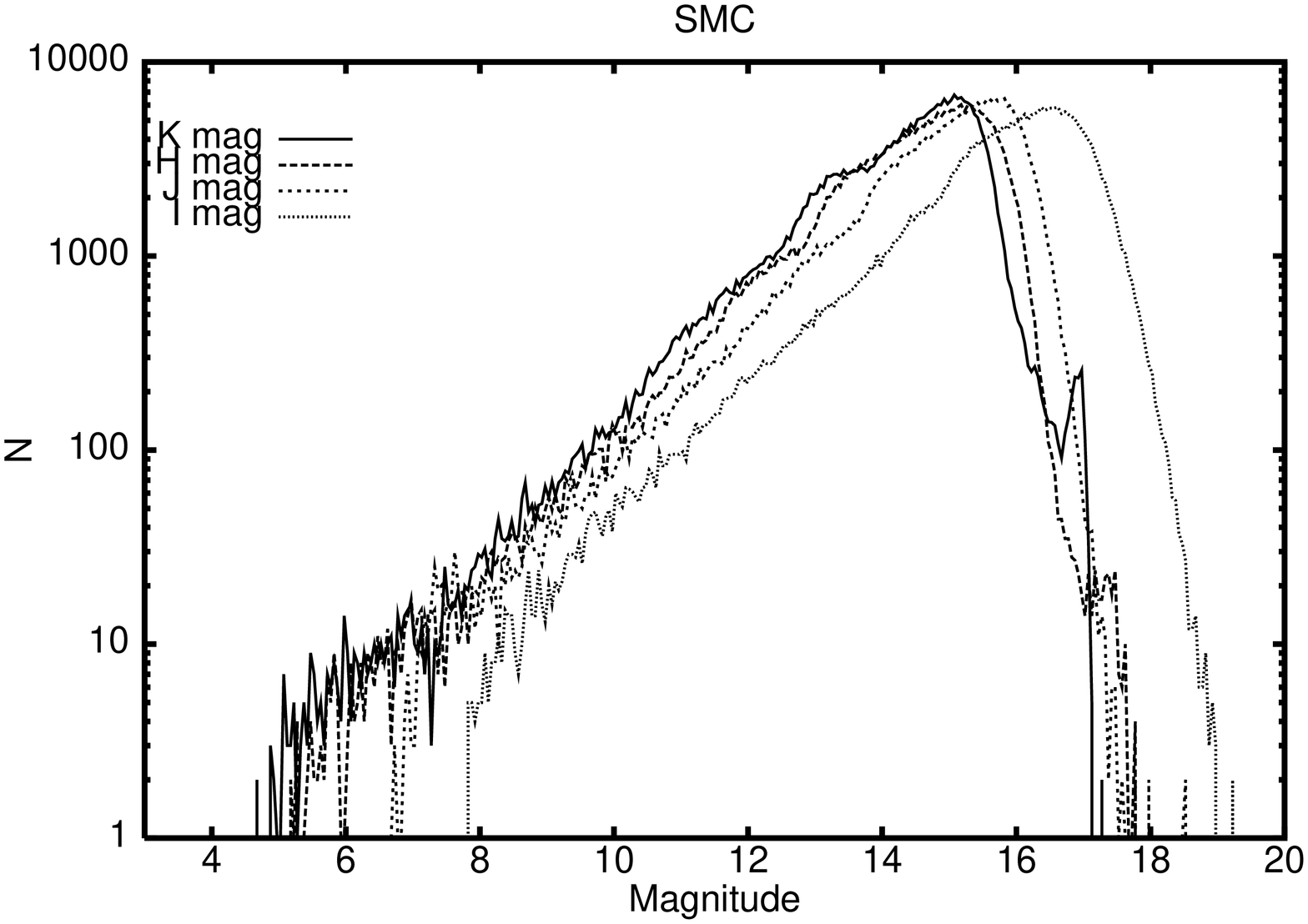,height=5cm,angle=-90} 
   \end{tabular}
   \end{center}
   \caption[example] 
   { \label{fig:completmc2}	  
Completeness diagrams for the merged DENIS and 2MASS point sources. } 
   \end{figure}

\begin{figure}
\begin{center}
   \begin{tabular}{c}
   \psfig{figure=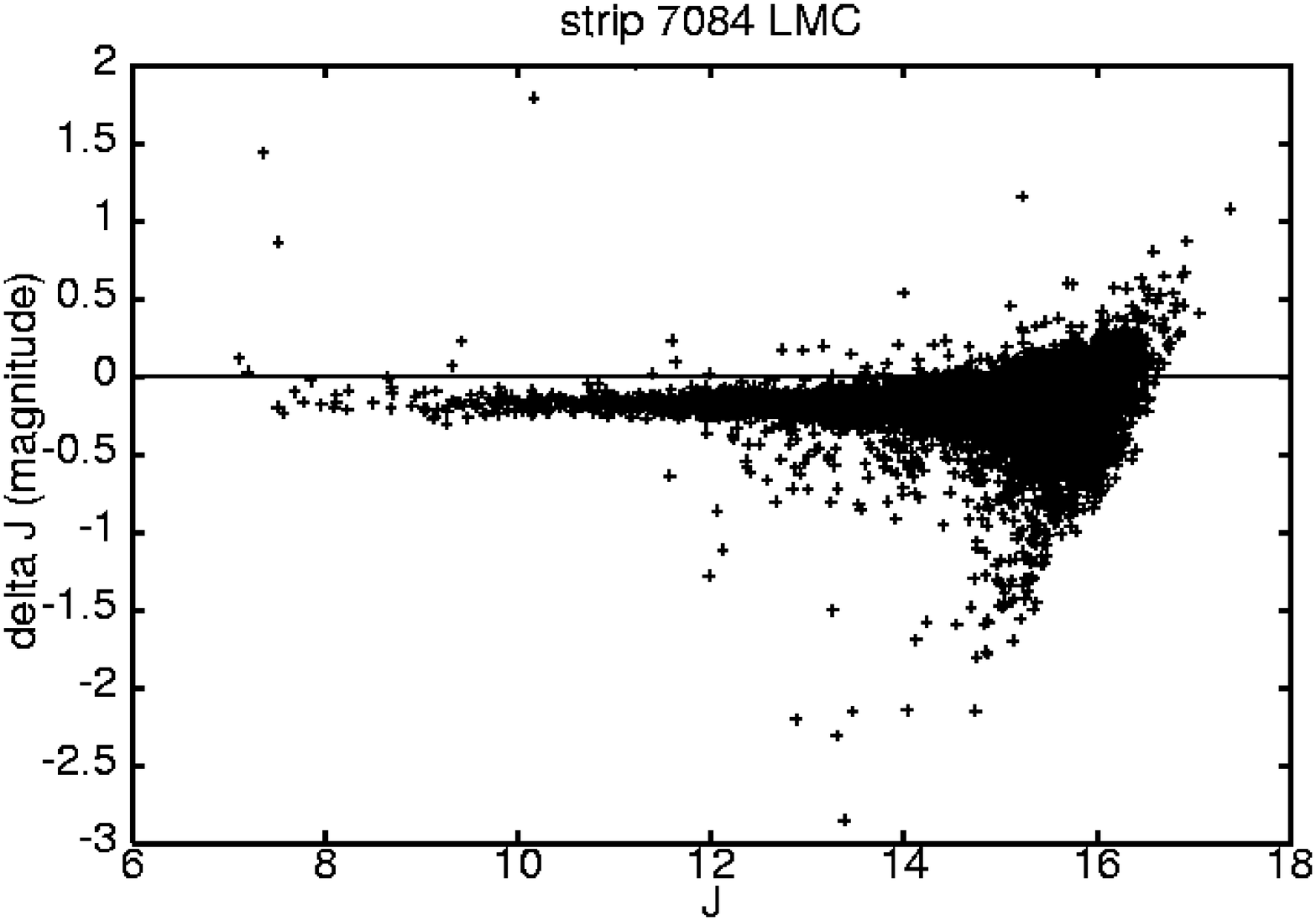,height=6cm} \psfig{figure=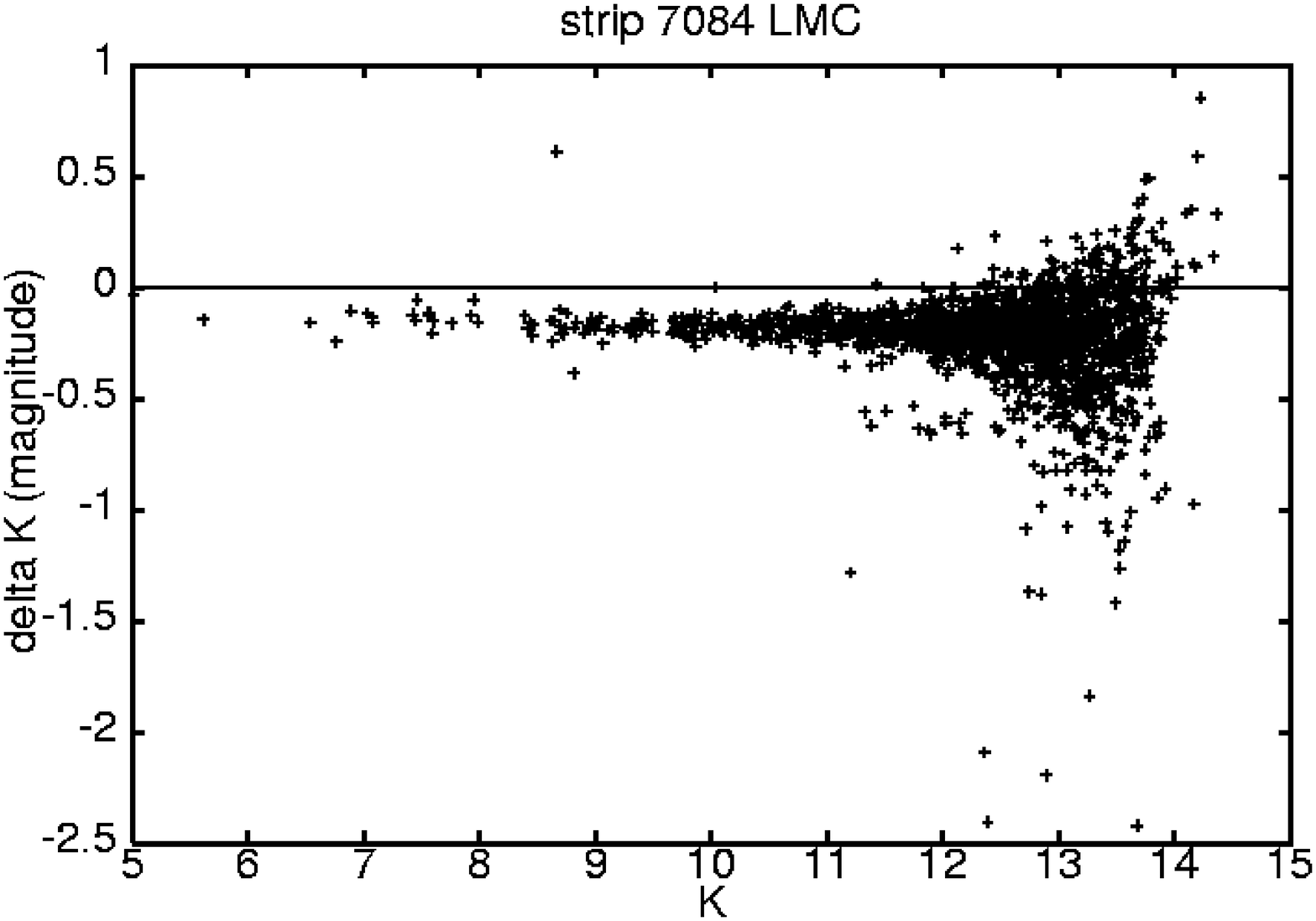,height=6cm} 
   \end{tabular}
   \end{center}
   \caption[example] 
   { \label{fig:biseau}	  
(Left)~$\textrm{J}_\textrm{DCMC} - \textrm{J}_\textrm{2MASS}$ as a function of $\textrm{J}_\textrm{DCMC}$. 
The shift is -0.182 for the range [10,14].
(Right)~$\textrm{K}_{\textrm{s}_\textrm{DCMC}} - \textrm{K}_{\textrm{s}_\textrm{2MASS}}$ as a function of $\textrm{K}_{\textrm{s}_\textrm{DCMC}}$.
The shift is -0.184 for the range [8,12]. } 
   \end{figure}

\begin{figure}
\begin{center}
   \begin{tabular}{c}
   \psfig{figure=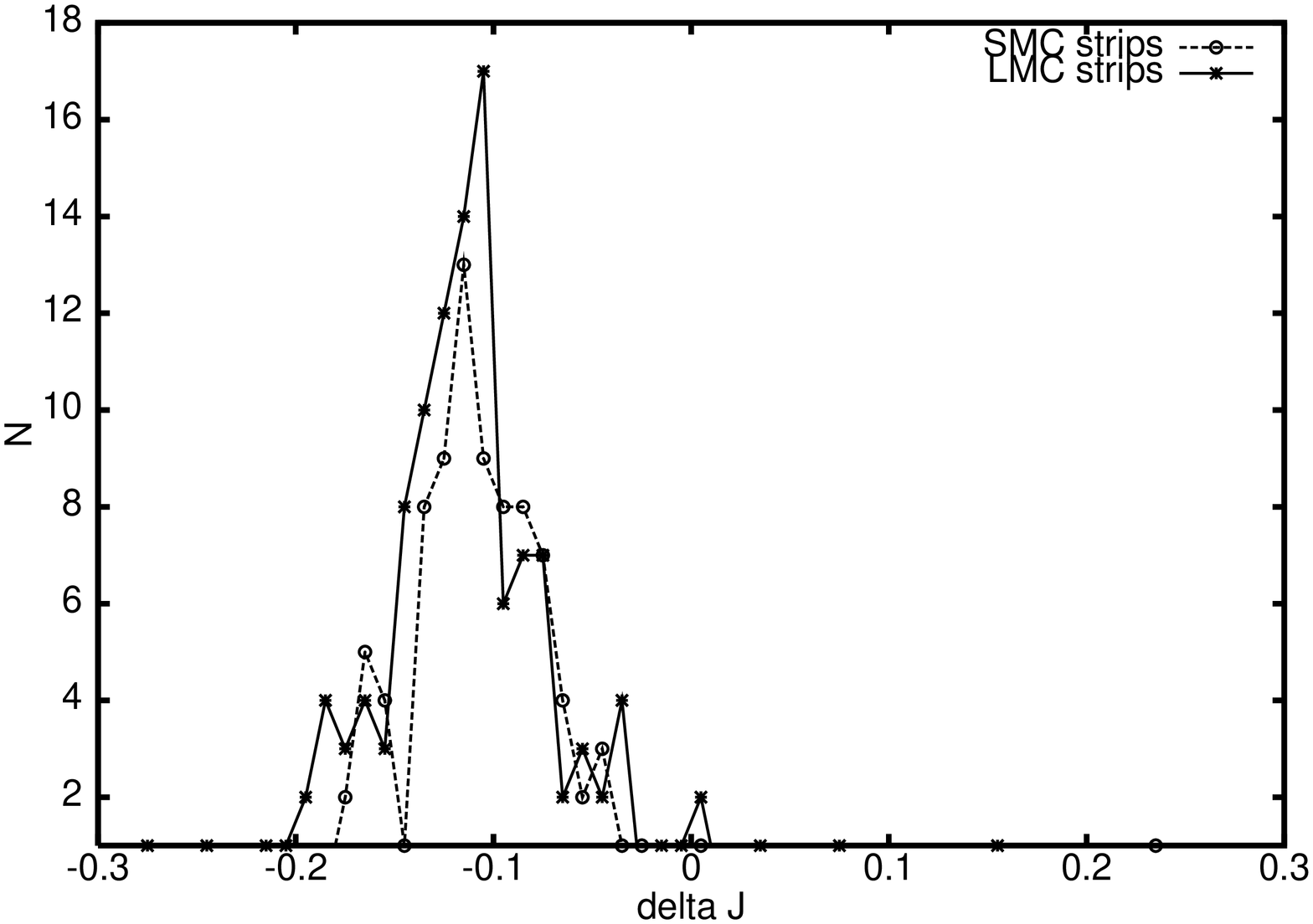,height=6cm,angle=-90} \psfig{figure=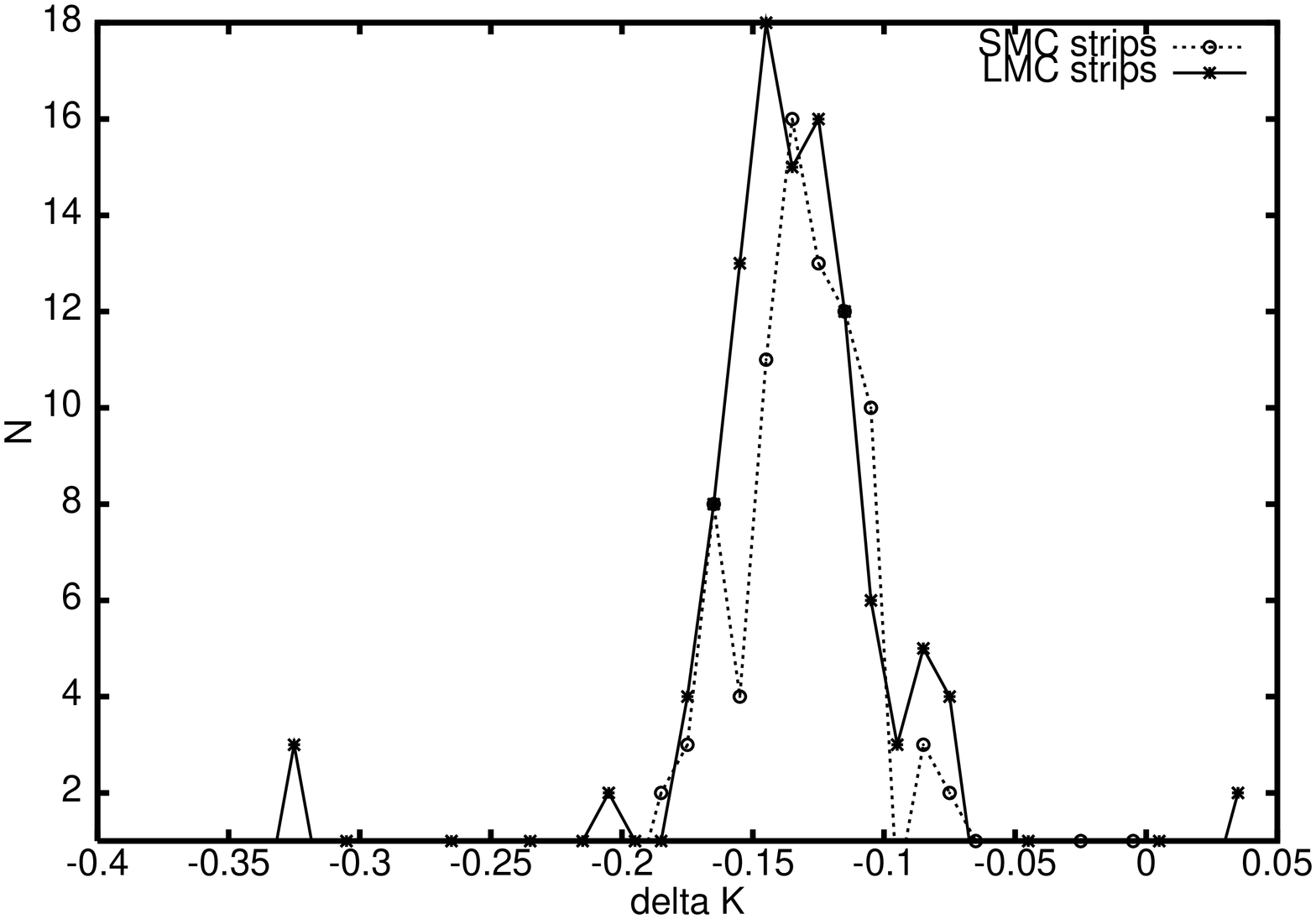,height=6cm,angle=-90}
   \end{tabular}
   \end{center}
   \caption[example] 
   { \label{fig:histodiff}	  
Histograms of the magnitude shifts found for the 119 LMC strips and 87 SMC strips.} 
   \end{figure}

     
We tried to find a mean relation between DCMC and 2MASS magnitudes, restricting to the range [10,14] in J
and [8,12] in $\textrm{K}_\textrm{s}$, avoiding saturated bright stars as well as the faintest ones.
There is a systematic shift of the absolute calibration between the two catalogues (Fig.~\ref{fig:biseau}). For each strip, we calculated the median 
of $\delta \textrm{J}$ and $\delta \textrm{K}_\textrm{s}$. Fig.~\ref{fig:histodiff} shows the histograms
of all the shifts found. The mean systematic shift between the two catalogues is -0.10 in J and -0.14 in $\textrm{K}_\textrm{s}$.

New color--magnitude diagrams (CMDs) produced out of the merged catalogue are presented as in Fig~11.

      \begin{figure}
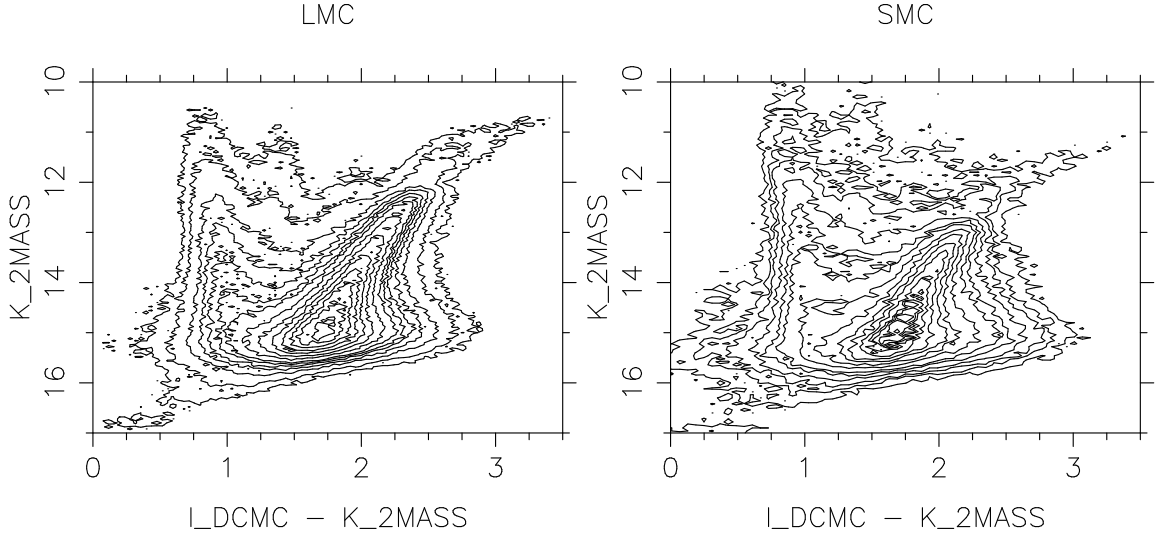

\begin{center}
   \begin{tabular}{c}
   \psfig{figure=HR_contour-LMC.ps,height=7cm,angle=-90} \psfig{figure=HR_contour-SMC.ps,height=7cm,angle=-90}
   \end{tabular}
   \end{center}
   \caption[example] 
   { \label{fig:contour}	  
CMDs for both Clouds (original in colors). There are 1251988 sources for the LMC and 278670 for the SMC.
We used the I band coming from the DCMC and the K band coming from 2MASS. 
Note that those observations 
  were not simultaneous so these CMDs 
 should be considered as indicative diagrams.} 
   \end{figure}

\section{CONCLUSION} 

The work presented here is an intermediary step before the production of a Master 
Catalogue of stars towards the Magellanic Clouds (MC2\cite{mc2}) which is to 
appear at the end of 2001. The Master Catalogue should also include cross--identifications 
with catalogues and tables at other wavelengths : GSC--II (optical), MSX
and IRAS (far infrared).\\
This reference catalogue will be made available as a support for a number of studies 
concerning, e.g. the stellar populations in the Magellanic Clouds, the structure of the 
Clouds, or certain classes of objects (Cepheids, AGB stars, etc.).  Recent articles, 
such as those  by Nikolaev \& Weinberg (2000)\cite{Nikolaev} and Cioni et al. (2000)\cite{2000A&A...358L...9C} have 
demonstrated the power of near infared surveys to improve our understanding of those 
neighbouring galaxies.

\acknowledgments     
This publication makes use of data products from the Two Micron All Sky Survey, which is
a joint project of the University of Massachusetts and the Infrared Processing and Analysis
Center/California Institute of Technology, funded by the National Aeronautics and Space
Administration and the National Science Foundation, and from DENIS, which is the result of a joint effort involving human and financial contributions of 
several Institutes mostly located in Europe. It has been supported financially mainly 
by the French Institut National des Sciences de l'Univers, CNRS, and French 
Education Ministry, the European Southern Observatory, the State of Baden-W\"urttemberg, and the European Commission under a network of the Human Capital 
and Mobility program.
%

  \bibliography{nozespie}   
  \bibliographystyle{spiebib}   
 
  \end{document}